\documentclass[10pt,aps,prc,twocolumn,superscriptaddress,floatfix]{revtex4-1}
\usepackage{graphicx}
\usepackage{amsmath}
\usepackage{braket}
\usepackage{url}
\usepackage{multirow}
\usepackage{amssymb} 
\usepackage{booktabs}
\usepackage{rotating}
\usepackage{bm}
\usepackage{bbm}
\usepackage{longtable}
\usepackage{subfigure}
\usepackage{tikz}
\usetikzlibrary{matrix}

\usepackage[colorlinks,
linkcolor=blue,
anchorcolor=red,
urlcolor=red,
citecolor=red]{hyperref}
\usepackage{cancel}

\usepackage[utf8]{inputenc}
\usepackage[T1]{fontenc}

\begin{document}

\title{\texorpdfstring{Large scale shell model calculation for collectivity in nuclei beyond $^{78}$Ni}{Large scale shell model calculation for collectivity in nuclei beyond 78Ni}}

\author{N. Chen}
\author{J. G. Li}\email[]{jianguo\_li@impcas.ac.cn}
\author{H. H. Li}
\affiliation{Heavy Ion Science and Technology Key Lab, Institute of Modern Physics, Chinese Academy of Sciences, Lanzhou 730000, China}
\affiliation{School of Nuclear Science and Technology, University of Chinese Academy of Sciences, Beijing 100049, China}

\date{\today}

\begin{abstract}

A  shell model effective interaction for nuclei beyond the double magic nucleus $^{78}$Ni is constructed. First, the single-particle evolutions for valence neutrons above the double magic $^{78}$Ni are systematically explored in the $N=51$ isotones using the large scale shell model (LSSM) calculations based on the constructed effective interaction.
Subsequently, we calculate the excitation energies of $2_1^+$ states and reduced electric quadrupole transition probabilities $B(E2;2^+ \to 0^+)$ for $N=52$ isotones. 
Notably, our calculation gives the result most consistent with the trend of the $B(E2)$ values observed in the $N=52$ isotones, especially for $^{84}$Ge, a result that poses a serious challenge to the theoretical model.
Furthermore, the collectivity in $N=52$ isotones, as well as the roles of pseudo-SU(3) symmetry, are investigated via the calculated primary configurations of their ground states and the first excited states.
Additionally, the low-lying structures and band characteristics of neutron-rich Ge and Se isotopes are investigated. The ground state and the $\gamma$-soft band are constructed in our LSSM calculations, aligning well with available experimental evidence. Finally, we present the calculated evolutions of low-lying states in neutron-rich Ge and Se isotopes. 
The predictions for the as-yet unobserved low-lying states in these nuclei provide a comprehensive dataset to guide and inform future experimental efforts to decipher the evolution of shell structures and collectivity.

\end{abstract}


\maketitle




\section{Introduction}

Nuclear dynamics of neutron-rich nuclei located far from the region of stability with a large $N/Z$ ratio have garnered considerable interest in recent years \cite{RevModPhys.83.1467,YANG2023104005,NOWACKI2021103866,PhysRevLett.124.222501,PhysRevC.95.041302,PhysRevLett.130.122502,PhysRevC.109.044313}.
A pivotal aspect of these systems is understanding the nuclear shape at low excitation energies, which plays a critical role in deciphering the complex interplay between collectivity and single-particle (s.p.) degrees of freedom within a nucleus \cite{RevModPhys.83.1467,YANG2023104005,NOWACKI2021103866,PhysRevC.95.041302,YUAN2024138331,PhysRevC.109.L041301}.
Nuclear collectivity has traditionally been thought to evolve from magic or semi-magic closed-shell nuclei characterized by seniority configurations through weakly deformed open-shell nuclei exhibiting vibrational characteristics and culminating in well-deformed nuclei with distinct rotational behavior \cite{RevModPhys.83.1467}.
The mass region near the robust $^{78}$Ni doubly magic nucleus is currently at the forefront of both experimental and theoretical research in nuclear structure \cite{78Ni-nature,PhysRevLett.116.182501,PhysRevLett.121.099902,PhysRevLett.117.172501,PhysRevLett.117.272501,PhysRevLett.116.182502,LI2023137893,PhysRevLett.124.222501,PhysRevC.95.041302}. 

The exploration of neutron-rich isotopes beyond $^{78}$Ni is poised to unveil new regions of nuclear deformation and phenomena of shape coexistence.
Intriguingly, nuclear deformation can emerge rapidly with the addition of just a few protons and neutrons.
In neutron-rich isotopes beyond $^{78}$Ni, nuclei transition from a spherical to a non-spherical shape, signifying a shift in their structural configuration.
By investigating the evolution of the nuclear deformation, particularly through the quadrupole correlations in isotopes and isotones beyond $^{78}$Ni, we can advance our comprehension of shell evolutions in exotic nuclei with extreme $N/Z$ ratios and provide a rich context for exploring s.p. motion, collective excitation, and nuclear deformations.
Recent experimental results have illuminated the low-lying levels in  $^{82,84}$Zn \cite{SHAND2017492}, suggesting the onset of deformation towards heavier Zn isotopes.
Moving further away from the $^{78}$Ni core, studies of Ge, As, and Se isotopes have revealed the emergence of $\gamma$ collectivity in isotopes with neutron numbers $N = 52$ and $N = 54$ \cite{PhysRevC.96.011301,PhysRevC.92.034305,PhysRevC.95.051302,PhysRevC.106.014320}.
Furthermore, the low-lying structure and shape evolution in neutron-rich $^{88,90,92,94}$Se isotopes have been comprehensively explored through in-beam $\gamma$-ray spectroscopy \cite{PhysRevC.95.041302}, revealing triaxiality and shape coexistence in these isotopes.
Notably, experimental results in $^{92}$Se and $^{94}$Se indicate a transition in ground-state shape from prolate deformation in $^{92}$Se to oblate in $^{94}$Se \cite{PhysRevLett.124.222501}. Moreover, an oblate $K$ isomer was also observed in $^{94}$Se \cite{PhysRevLett.124.222501}. 
Furthermore, the mass region surrounding $^{78}$Ni is not only pivotal for nuclear structure studies but also plays a crucial role in our understanding of the $r$-process nucleosynthesis~\cite{PhysRevLett.101.262501,PhysRevLett.110.041101,PhysRevLett.109.112501}.

The large scale shell model (LSSM) has enabled enormous progress in investigating the neutron-rich nuclei above $^{78}$Ni. 
Nuclei beyond $^{78}$Ni can be effectively described within a model space comprising protons in the \{$0f_{5/2},1p_{1/2,3/2},0g_{9/2}$\} orbitals and neutrons in the \{$1d_{3/2,5/2},2s_{1/2},0g_{7/2},0h_{11/2}$\} orbitals. Notably, in nuclei situated beyond $^{78}$Ni, the proton $0f_{5/2}$ and $1p_{3/2}$ s.p. states, as well as the neutron $1d_{5/2}$ and $2s_{1/2}$ s.p. states, exhibit near degeneracy. These states form the foundation blocks of the \textit{pseudo}-SU(3) and \textit{quasi}-SU(3) symmetries, respectively, as detailed in Refs.~\cite{RevModPhys.77.427,PhysRevC.88.034327,PhysRevLett.121.192502,LI2023137893}. Those symmetries plays a crucial role in the formation of nuclear deformations, making the nuclei situated above $^{78}$Ni particularly interesting for studying pseudospin symmetry, nuclear deformation and SU(3) symmetries~\cite{RevModPhys.77.427,PhysRevLett.121.192502}. 
In addition, the effective shell model interaction jj45, constructed above the $^{78}$Ni inner core \cite{PhysRevLett.91.162503}, has provided insights. It is primarily tailored for nuclei southwest of $^{132}$Sn and not optimally suited for those near $^{78}$Ni. To address this gap, Sieja \textit{et al.} have a developed shell model interaction for nuclei above $^{78}$Ni~\cite{PhysRevC.79.064310,PhysRevC.88.034327}. The interaction has successfully described low-lying states for nuclei above $^{78}$Ni, such as the vibrational $\gamma$-soft band in $^{84,86,88}$Ge \cite{PhysRevC.96.011301} and $^{86,88}$Se \cite{PhysRevC.92.034305,PhysRevC.95.051302}, as well as the structures of even-even $^{92,94,96}$Sr isotopes \cite{PhysRevC.104.064309}. However, the effective shell model interaction has shown limitations in accurately describing the low-lying states of nuclei close to $^{78}$Ni \cite{SHAND2017492}, as well as in the calculation of reduced electric quadrupole transition probabilities $B(E2;2^+ \to 0^+)$ in the $N=52$ isotones \cite{PhysRevLett.121.192502}.
Recently, two new shell model interactions have been developed that are suitable for calculating levels in isotopes near $^{78}$Ni within cross-shell model spaces \cite{LI2023137893,PhysRevLett.117.272501}.
However, the computational costs for the nuclei above $^{78}$Ni using these interactions are extremely large and beyond the capability of supercomputers.
Additionally, an effective interaction, named DF2882, has been successfully employed to describe the neutron-rich As isotopes, with the results being accurately analyzed in terms of pseudo-SU3 symmetry~\cite{PhysRevC.106.014320}.
Moreover, the realistic effective shell model has also been constructed to calculate the double $\beta$ decay of $^{100}$Mo~\cite{PhysRevC.105.034312,PhysRevC.91.041301}, as well as the nuclear structures in Zr, Mo, Ru, Pd, and Sn isotopes located above $^{78}$Ni~\cite{PhysRevC.93.064328}.
Furthermore, the interacting boson model \cite{PhysRevC.91.054304,PhysRevC.105.054318} and self-consistent beyond-mean-field calculations~\cite{PhysRevC.95.041302} have also been applied to investigate the level structure and electromagnetic transitions in these neutron-rich nuclei \cite{PhysRevC.91.054304,PhysRevC.105.054318}.







In this study, we undertake comprehensive shell model calculations for the nuclei situated above the doubly magic $^{78}$Ni, utilizing an effective interaction specifically constructed for this investigation.
The article is structured as follows: a brief introduction of the effective shell model interaction used in the present work is outlined in the Method section. Subsequent sections are dedicated to systematic large scale shell model calculations. We explore the evolution of energies of states with single-particle nature in $N=51$ isotones and examine the excitation energies [$E(2_1^+)$] and electric quadrupole transition probabilities [$B(E2;2_1^+ \to 0_1^+)$] for $N=52$ isotones. Additionally, we delve into the collective behaviors exhibited by the ground states and $\gamma$-soft bands, as well as the evolution of low-lying states in even-even neutron-rich Ge and Se isotopes.


\section{Method}

The shell model calculations begin with the effective interaction $V_{\rm eff}$. It is based on the definition of a restricted valence space, where a suitable Hamiltonian could be diagonalized. In recent years, several many-body methods have been employed to derive the effective shell model interaction, such as open-shell many-body perturbation theory (MBPT) \cite{HJORTHJENSEN1995125,CORAGGIO20122125,PhysRevC.102.034302}, the valence-space in-medium similarity renormalization group \cite{HERGERT2016165,doi:10.1146/annurev-nucl-101917-021120,PhysRevLett.106.222502,PhysRevLett.118.032502,PhysRevLett.113.142501,PhysRevC.107.014302,LI2023138197,CPC:10.1088/1674-1137/acf035}, the coupled-cluster shell model \cite{PhysRevC.98.054320,PhysRevC.104.064310}, and the no-core shell model with a core \cite{PhysRevC.78.044302}. 
As done in Ref.~\cite{PhysRevC.105.034312}, the MBPT calculation based on realistic charge-dependant Bonn potential (CD-Bonn) nucleon-nucleon interaction~\cite{PhysRevC.63.024001} is adopted to construct the effective interaction in the present work.
To remove the strong short-range repulsive core in the bare CD-Bonn interaction, the low-momentum approach $V_{\rm{low}-k}$ \cite{PhysRevC.65.051301,HJORTHJENSEN1995125,PhysRevC.105.034312} is applied, resulting in a more manageable potential for nuclear structure calculations.
This advanced approach has been proven to be quite successful in describing the spectroscopic properties of nuclei from various regions \cite{PhysRevC.95.021304,PhysRevC.96.034312,PhysRevC.60.064306,LI2023137893,PhysRevC.102.034302,MA2020135257,PhysRevC.89.024319,PhysRevC.90.024312,PhysRevC.108.L031301,PhysRevC.105.034312}.
In the present calculations, a smooth low momentum potential is obtained by integrating $V_{NN}$ to a cutoff momentum $\Lambda = 2.6$ fm$^{-1}$. 
Within the open-shell MBPT derivation for the effective interaction, all the $\hat{Q}$-box vertex functions up to the third order of Goldstone diagrams \cite{HJORTHJENSEN1995125,CORAGGIO20122125} are included, and for a given $\hat{Q}$-box, the folded diagram series is summed up to all orders \cite{HJORTHJENSEN1995125,CORAGGIO20122125}. 
For our purposes, an effective interaction is derived in a valence space where valence protons and neutrons are active in the $\{1p_{3/2,1/2}, 0f_{5/2},0g_{9/2}\}$ and $\{1d_{5/2,3/2}$, $2s_{1/2}$, $0g_{7/2}$, $0h_{11/2}\}$ spaces, above the $^{78}$Ni inner core, respectively.

\begin{center}
\begin{table}[!htb]
    \centering
    \setlength{\tabcolsep}{1.0mm}
    \begin{tabular}{cccccccc}
    \hline\hline
     proton orbitals & this paper & Ref.~\cite{PhysRevC.105.034312} & Ref.~\cite{PhysRevC.79.064310} \\
    \hline

    $\pi0f_{5/2}$ & 0.00 & 0.00 & 0.27 \\
    $\pi1p_{3/2}$ & 0.65 & 1.60 & 0.00 \\
    $\pi1p_{1/2}$ & 2.21 & 2.10 & 2.26 \\
    $\pi0g_{9/2}$ & 3.00 & 4.30 & 3.89 \\

      \hline
      neutron orbitals &  this paper & Ref.~\cite{PhysRevC.105.034312} & Ref.~\cite{PhysRevC.79.064310} \\
      \hline
      $\nu1d_{5/2}$ & 0.00 & 0.40 & 0.00 \\
      $\nu2s_{1/2}$ & 0.28 & 0.00 & 0.00 \\
      $\nu1d_{3/2}$ & 1.55 & 1.10 & 1.99 \\
      $\nu0g_{7/2}$ & 1.80 & 2.80 & 1.99 \\
      $\nu0h_{11/2}$ & 3.80 & 3.20 & 4.39\\
    \hline\hline
    \end{tabular}
    \caption{Proton and neutron s.p. energies (in MeV) used in our work, along with the values adopted in Refs.~\cite{PhysRevC.105.034312,PhysRevC.79.064310} for comparison.}
    \label{tab}
\end{table}
\end{center}

Monopole matrix elements were adjusted in the effective interaction derived from the realistic CD-Bonn interaction to produce the low-lying states of Ge, As, and Se isotopes with $N = 50, 51$, and 52, in which the effects of three-body force can be considered within the optimization~\cite{PhysRevLett.90.042502}.
In the real calculations, adjustments are made for the proton-proton monopole matrix elements between $\pi 0f_{5/2}$ and $\pi 1p_{1/2}$ orbitals, neutron-neutron monopole matrix elements between $\nu 1d_{5/2,3/2}$ and $\nu 2s_{1/2}$ orbitals, and proton-neutron monopole matrix elements between proton $\pi 0f_{5/2}$ and $\pi 1p_{1/2}$ orbitals and neutron $\nu 1d_{5/2}$ and $\nu 2s_{1/2}$ orbitals.
The single-particle energies of orbitals within model space in our works are presented in Table~\ref{tab}. Additionally, for comparison, we also include the single-particle energies used in the shell model interactions with the same model space as reported in Refs.~\cite{PhysRevC.105.034312,PhysRevC.79.064310}.


Finally, the optimized shell model effective Hamiltonian is exactly diagonalized using the KSHELL shell model code \cite{shimizu_thick-restart_2019}.
An enhanced polarization charge of $0.7e$ is used for calculations involving quadrupole moments and transition rates, as previously suggested for this model space \cite{PhysRevC.88.034327,PhysRevC.79.064310}. In the present work, most of the nuclei are calculated without truncation, except that the $4\hbar w$ truncations are adopted for $^{88}$Ge and $^{90}$Se, and $2\hbar w$ truncations are used for $^{90,92}$Ge and $^{92,94}$Se. The largest Hamiltonian dimension is about $1.0\times 10^{9.22}$.

\section{Results}

\begin{figure}[!htb]
\includegraphics[width=0.45\textwidth]{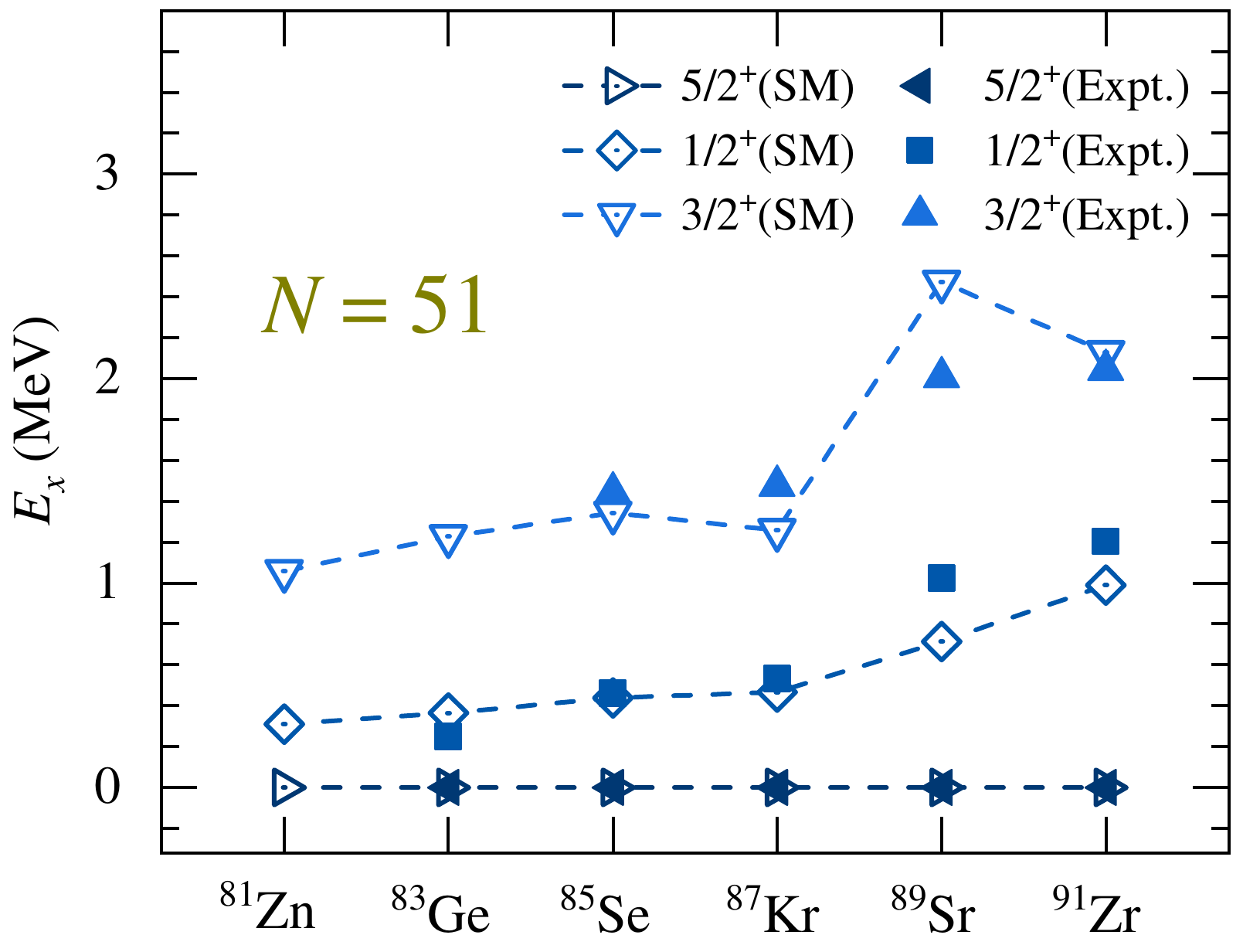}
\caption{\label{N=51} Calculated excitation energy of $1/2^+$ and $3/2^+$ states for the light $N=51$ odd isotones from $Z=30$ to 40 using LSSM, along with available experimental data \cite{ensdf}.}
\end{figure}

While the $N = 50$ shell-gap evolution towards $^{78}$Ni has been and continues to be the object of intense experimental investigations worldwide, attempts to determine the energies of states with single-particle nature above the $^{78}$Ni core remain scarce. 
Detailed information on the states with single-neutron properties above $N = 50$ is available close to stability \cite{PhysRevC.87.014312}. However, their evolutions towards $Z = 28$ are still poorly known or understood both from theoretical and experimental points of view. 
Such efforts for precise calculations using the model space above $^{78}$Ni to follow these evolutions are not only necessary but would significantly contribute to the predictive power required for designing and interpreting future experiments.

The systematics of the excitation energies of low-lying $1/2^+$ and $3/2^+$ states in $N=51$ odd isotones are calculated, and the results are presented in Fig.~\ref{N=51}, alongside experimental data for comparison \cite{ensdf}.
It is well established that the ground states of the $N = 51$ odd isotones, ranging from $^{83}$Ge to $^{91}$Zr, are $5/2^+$, stemming from the occupation of the $\nu 1d_{5/2}$ orbital by a single valence neutron. Our calculations give $1/2^+$ as the first excited state, in agreement with the experimental data for all the isotones considered. The evolution of the $1/2^+$ states shows a continuous decrease from stability towards $Z = 30$.
The general trend of the calculated $3/2^+$ states aligns closely with the experimental results for all isotones, except for the state in $^{89}$Sr, where the theoretical calculations are marginally higher. The trend remains relatively flat from $Z = 30$ to 36, while at $^{89}$Sr, there is a notable increase in the energies of the $1/2^+$ and $3/2^+$ states, a pattern also observed in $^{91}$Zr. This abrupt surge may be indicative of changes in shell filling and wave function components at $Z = 38$ and 40.

\begin{figure}[!htb]
\includegraphics[width=0.42\textwidth]{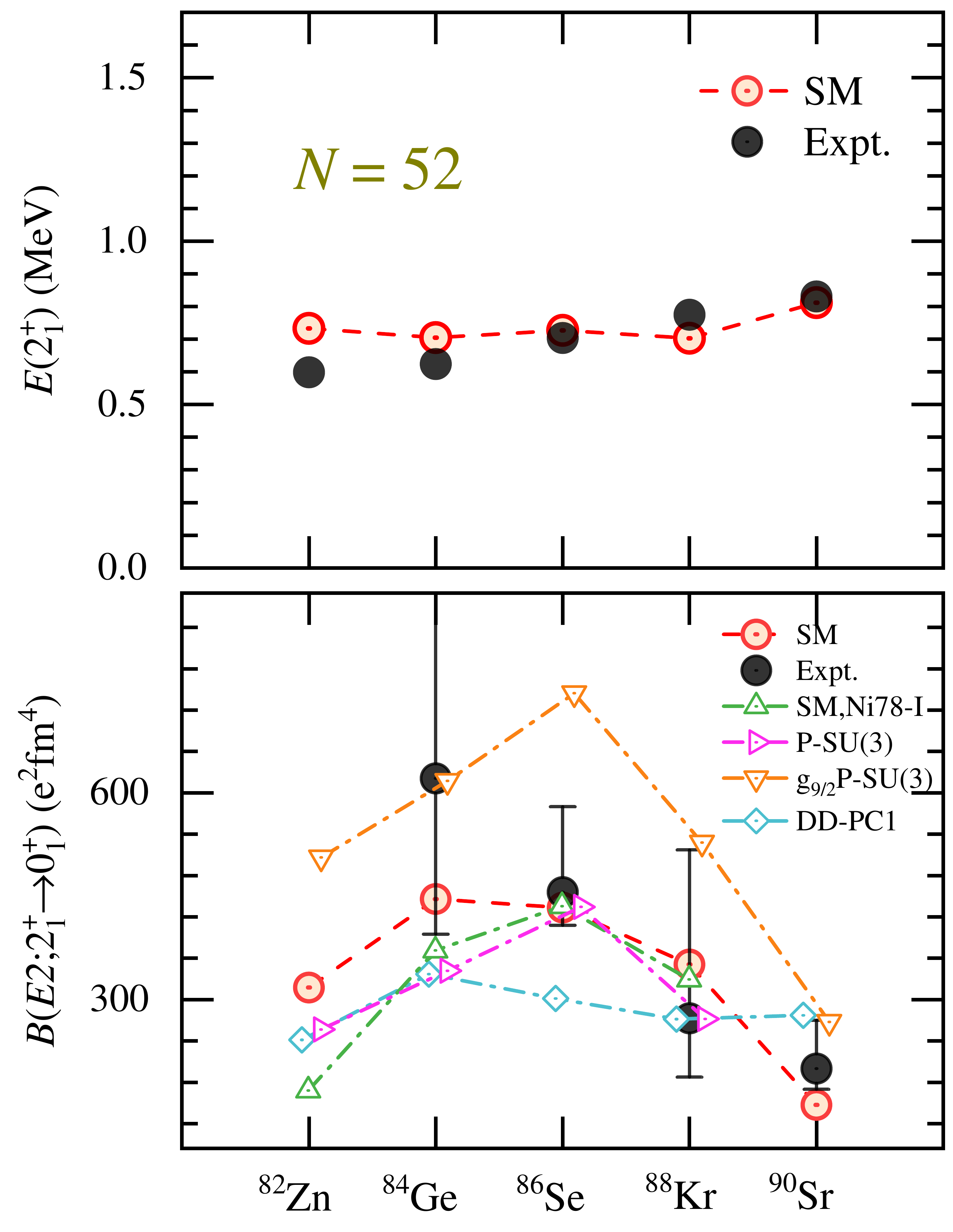}
\caption{\label{N=52}Calculated excitation energy of $2_1^+$ and $B(E2;2^+\to0^+)$ systematics of the $N = 52$ even-even isotones from $Z = 30$ to $Z = 38$ with LSSM, along with experimental data \cite{PhysRevLett.121.192502} and theoretical calculations presented in Ref.~\cite{PhysRevLett.121.192502}.}
\end{figure}

Recently, lifetime measurements of excited states in the light $N=52$ isotones $^{88}$Kr, $^{86}$Se, and $^{84}$Ge have been performed \cite{PhysRevLett.121.192502}, yielding the reduced electric quadrupole transition probabilities $B(E2; 2^+\to0^+)$ of these nuclei. 
The results indicate that the maximum quadrupole collectivity occurs in $^{84}$Ge within the $N=52$ isotones, ranging from $^{90}$Sr to $^{84}$Ge. 
Although the obtained $B(E2; 2^+\to0^+)$ of $^{84}$Ge bears large uncertainties, the trend of $B(E2; 2^+\to0^+)$ in $N=52$ isotones still provide great challenges for theoretical calculations. We have also performed LSSM calculations based on the effective interaction constructed in this work for the $N=52$ isotones. The calculated excitation energies of $2_1^+$ [$E(2_1^+)$] and $B(E2; 2^+\to0^+)$ values are presented in Fig.~\ref{N=52}. These are compared with experimental data and theoretical results from shell model calculations using the Ni78-I interaction, the pseudo-SU(3) limit [P-SU(3)], the pseudo-SU(3) limit including one $N=50$ core-breaking $0g_{9/2}$ pair promotion [$g_{9/2}$P-SU(3)], and beyond mean-field calculations using the relativistic functional DD-PC1.



It is evident that the calculated $E(2^+_1)$ values agree well with the experimental data, exhibiting a flat and stable trend across the change in proton number, with the maximum discrepancy being a mere $0.15$ MeV for $^{82}$Zn.

The states with single-particle nature in the $N=51$ isotones above $^{78}$Ni contribute to the formation of pseudo-SU(3) blocks for protons, involving the $1p_{3/2}$ and $0f_{5/2}$ orbits. LSSM calculations of $B(E2;2^+ \to 0^+)$ using the Ni78-I effective interaction are in agreement with experimental data for $^{88}$Kr and $^{86}$Se, as are the pseudo-SU(3) calculations. However, both models suggest the largest collectivity in $^{86}$Se, with the $B(E2;2^+ \to 0^+)$ values for $^{84}$Ge still smaller than the lower bounds of the experimental uncertainty.
Furthermore, calculations using the $g_{9/2}$P-SU(3) model accurately reproduce the experimental $B(E2;2^+ \to 0^+)$ values for $^{84}$Ge. Unfortunately, the results of this model significantly diverge from the experimental results for other $N=52$ isotones. In addition, the results from the relativistic functional DD-PC1 present an almost perfect linear relationship for $B(E2;2^+ \to 0^+)$ across $N=52$ isotones, with $^{82}$Zn being the sole exception.



\begin{table}[!htb]
    \centering
    \setlength{\tabcolsep}{1.0mm}
    \begin{tabular}{cccc}
    \hline\hline
    Nuclei & $J^\pi$ & Configuration & Prob. \\
    \hline
    $^{82}$Zn & $0^+$ & $\pi(1p_{3/2})^2\otimes\nu(1d_{5/2})^2$ & 30\%\\
     & & $\pi(0f_{5/2})^2\otimes\nu(1d_{5/2})^2$ & 23\%\\
     & $2^+$ & $\pi(1p_{3/2})^2\otimes\nu(1d_{5/2})^2$ & 23\%\\
     & &  $\pi(0f_{5/2})^2\otimes\nu(1d_{5/2})^2$ & 17\%\\
     & & $\pi(1p_{3/2})^2\otimes\nu\{(2s_{1/2})^1(1d_{5/2})^1\}$ & 11\%\\
    $^{84}$Ge & $0^+$ & $\pi\{(1p_{3/2})^2(0f_{5/2})^2\}\otimes\nu(1d_{5/2})^2$ & 42\%\\
     & $2^+$ & $\pi\{(1p_{3/2})^2(0f_{5/2})^2\}\otimes\nu(1d_{5/2})^2$ & 38\%\\
      & & $\pi\{(1p_{3/2})^2(0f_{5/2})^2\}\otimes\nu\{(2s_{1/2})^1(1d_{5/2})^1\}$ & 14\%\\
    $^{86}$Se & $0^+$ & $\pi\{(1p_{3/2})^4(0f_{5/2})^2\}\otimes\nu(1d_{5/2})^2$ & 28\%\\
     & & $\pi\{(1p_{3/2})^2(0f_{5/2})^4\}\otimes\nu(1d_{5/2})^2$ & 26\%\\
      & $2^+$ & $\pi\{(1p_{3/2})^2(0f_{5/2})^4\}\otimes\nu(1d_{5/2})^2$ & 25\%\\
     & & $\pi\{(1p_{3/2})^4(0f_{5/2})^2\}\otimes\nu(1d_{5/2})^2$ & 23\%\\
    $^{88}$Kr & $0^+$ & $\pi\{(1p_{3/2})^4(0f_{5/2})^4\}\otimes\nu(1d_{5/2})^2$ & 50\%\\
     & $2^+$ & $\pi\{(1p_{3/2})^4(0f_{5/2})^4\}\otimes\nu(1d_{5/2})^2$ & 49\%\\
     & & $\pi\{(1p_{3/2})^4(0f_{5/2})^4\}\otimes\nu\{(2s_{1/2})^1(1d_{5/2})^1\}$ & 10\%\\
     $^{90}$Sr & $0^+$ & $\pi\{(1p_{3/2})^4(0f_{5/2})^6\}\otimes\nu(1d_{5/2})^2$ & 59\%\\
      & $2^+$ & $\pi\{(1p_{3/2})^4(0f_{5/2})^6\}\otimes\nu(1d_{5/2})^2$ & 60\%\\ 
    \hline\hline
    \end{tabular}
    \caption{The contributions of the configurations in the ground states and $2^+$ excited states of $N=52$ isotones. Only the contributions larger than 10\% are presented.}
    \label{tab:configuration}
\end{table}

While the experimental error bars for the $B(E2)$ values of  $^{84}$Ge are large, the results of the various theoretical calculations discussed above are still unsatisfactory and have limitations. 
The calculated $B(E2)$ values of $^{84}$Ge in our work are located within the errors of experimental data. Moreover, the general trend of the $B(E2)$ values calculated in our work is closer to the experimental results for all these isotones than the results from other approaches; see details in Fig.~\ref{N=52}.
It is noteworthy that the largest value of $B(E2;2^+ \to 0^+)$ is found at $^{84}$Ge in our shell model calculations, which is also the case experimentally~\cite{PhysRevLett.121.192502}, suggesting that the maximum deformation in the $N=52$ isotope occurs at $^{84}$Ge.
We also calculate the primary configurations of the ground states and the first excited states for $N=52$ isotones which are listed in Table~\ref{tab:configuration}. Our LSSM calculations show that the proton configurations for both the ground states and the first excited states $2^+$ for $N=52$ isotones ranging from $Z = 32$ to $38$ are predominantly a mix of the $1p_{3/2}$ and $0f_{5/2}$ orbits, reflecting a clear manifestation of pseud-SU(3) symmetry~\cite{PhysRevLett.78.436}. Moreover, the near-degeneracy or close proximity of the $1p_{3/2}$ and $0f_{5/2}$ orbitals approaches the SU(3) limit, which contributes significantly to the quadrupole character of the nucleus. The emergence of maximum collectivity in $^{84}$Ge among the $N=52$ isotones is interpreted as a result of a pseudo-SU(3) structure, in which $^{84}$Ge is located in the middle of the $1p_{3/2}$ and $0f_{5/2}$ orbitals for the $N=52$ isotones. For the neutron part, the nucleons largely occupy the same $1d_{5/2}$ orbit. Furthermore, significant contributions to the collectivity may arise from the configuration mixing between the \textit{quasi}-SU(3) partner orbits $1d_{5/2}$ and $2s_{1/2}$ \cite{PhysRevC.92.024320}.

\begin{figure*}[!htb]
\centering
\includegraphics[width=1\textwidth]{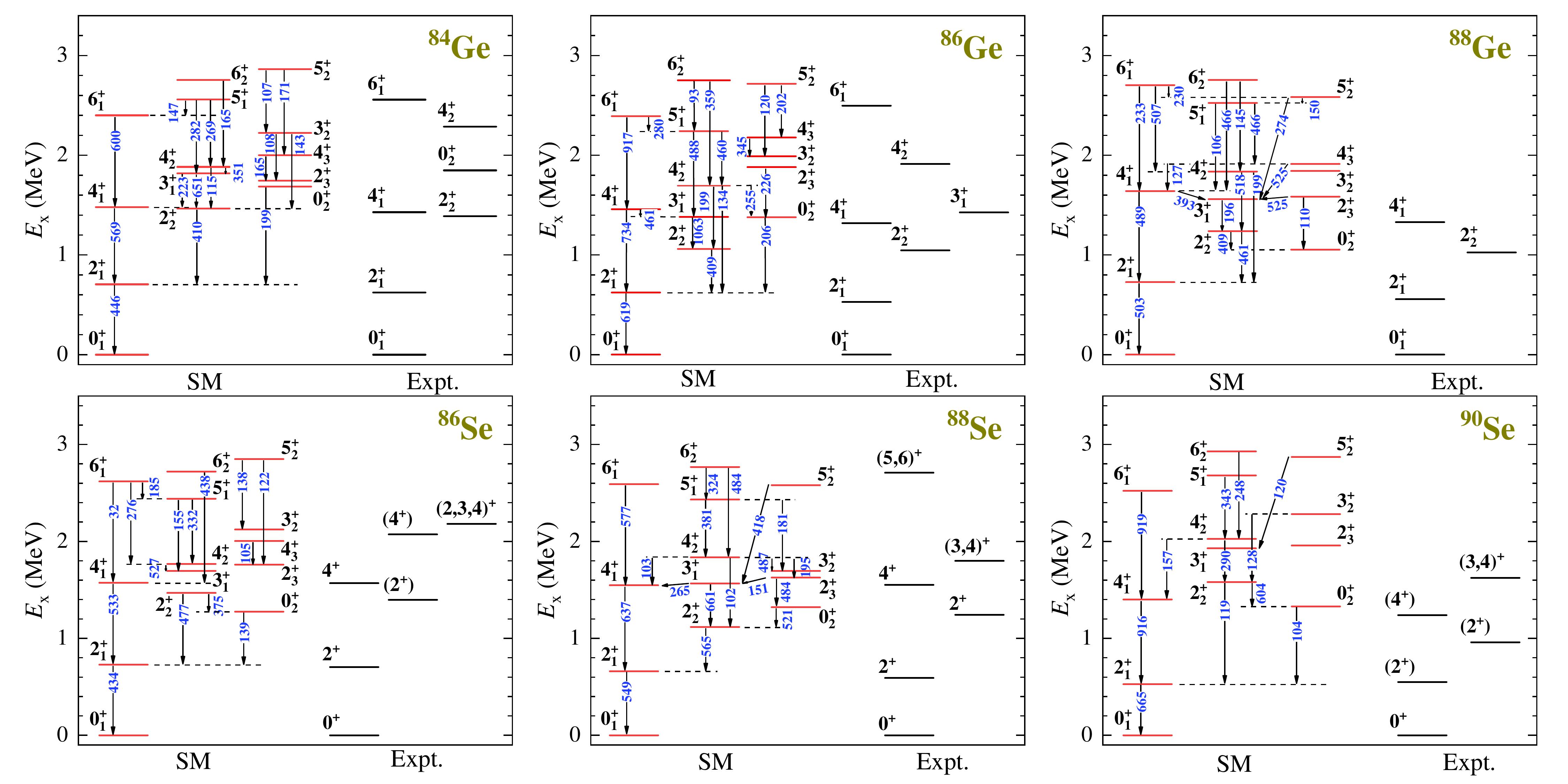}
\caption{\label{level scheme}  {Excitation energies and reduced transition rates $B(E2)$ (in ${\rm e^2fm^4}$) for Ge and Se isotopes, calculated in the shell model framework, along with the experimental data for comparison \cite{ensdf}.}}
\end{figure*}

Exploring the transition from closed-shell to open-shell nuclei reveals shifts from spherical to deformed shapes along isotopic and isotonic chains.
Notably, experimental evidence indicates that starting from the spherical $^{82}$Ge with $N = 50$, neutron-rich Ge isotopes evolve towards a potential region of rigid triaxial deformation in $^{86,88}$Ge \cite{PhysRevC.80.044308,PhysRevC.96.011301}. 
A similar transformation is observed in the Se isotopes \cite{PhysRevC.92.034305,PhysRevC.92.064322}. 
Consequently, the predicted emergence of a triaxial deformation region within neutron-rich Ge and Se isotopes demands further comprehensive theoretical exploration.
We have calculated the level structure and $B(E2)$ transitions between the low-lying states in neutron-rich $^{84,86,88}$Ge and $^{86,88,90}$Se isotopes. The results are presented in Fig. \ref{level scheme}, with the calculated $B(E2)$ values utilized to construct the band structure of these isotopes.

Our shell model calculations reasonably well reproduce the general features of the experimental level scheme for $^{84,86,88}$Ge and $^{86,88,90}$Se.
These nuclei exhibit low $2_1^+$ excitation energies and large $B(E2; 2^+_1 \to 0^+_1)$ values, indicative of significant deformations.
Additionally, a low-lying $2^+_2$ state, found below the $4_1^+$ states in the low-excitation region, suggests that these nuclei may exhibit triaxiality or $\gamma$-softness properties.
An exception is the $2_2^+$ state in $^{90}$Se, where our shell model calculations yield the excitation energy higher than the experimental data, placing it above the $4_1^+$ states.
Significantly, large $B(E2; 2^+_2\to2^+_1)$ values, comparable to the $B(E2; 2^+_1 \to 0^+_1)$ transitions, are observed in these isotopes, hinting at the possibility of collective vibrational modes.
Furthermore, the presence of a $0_2^+$ state in the low-lying region in our LSSM suggests potential shape coexistence in these nuclei, though further experimental investigation of the $B(E0)$ transition between the two $0^+$ states is required.
The $B(E2;0^+_2 \to 2_1^+)$ and $B(E2;2_2^+ \to 0_2^+)$ transitions in $^{84,86,88}$Ge and $^{86,88,90}$Se are generally large in our LSSM calculations. Exceptions include smaller $B(E2;2_2^+ \to 0_2^+)$ transitions in $^{84,86}$Ge and smaller $B(E2;0^+_2 \to 2_1^+)$ in $^{88}$Se, indicating less significant transitions between these states.
These calculations collectively indicate significant deformations in these nuclei.

Additionally, the band structures in the low-lying states of these nuclei are constructed based on the calculated $B(E2)$ values, where large $E2$ transitions occur between states within the same band.
Our LSSM calculations clearly indicate that the ground state band, comprising the yrast $0_1^+$, $2_1^+$, $4_1^+$, and $6_1^+$ states, is well defined in the $^{84,86,88}$Ge and $^{86,88,90}$Se isotopes, and exhibits large $B(E2)$ values among the states within the bands.
Significant $B(E2)$ transitions are observed between the states within the ground state band.
Adjacent to the ground state band, significant $B(E2)$ transitions exist between the $2_2^+$, $3_1^+$, $4_2^+$, $5_1^+$, and $6_2^+$ states in these nuclei, forming a $\gamma$-soft band above the $2^+_2$ state.
The presence of the $\gamma$-soft band in these nuclei further supports their exhibition of triaxiality properties.
Importantly, strong $E2$ transitions are also observed between states in the ground state band and the $\gamma$-soft band, which supports the collective nature of low-lying spectra in $^{84,86,88}$Ge and $^{86,88,90}$Se isotopes.

\begin{figure}[!htb]
\includegraphics[width=0.48\textwidth]{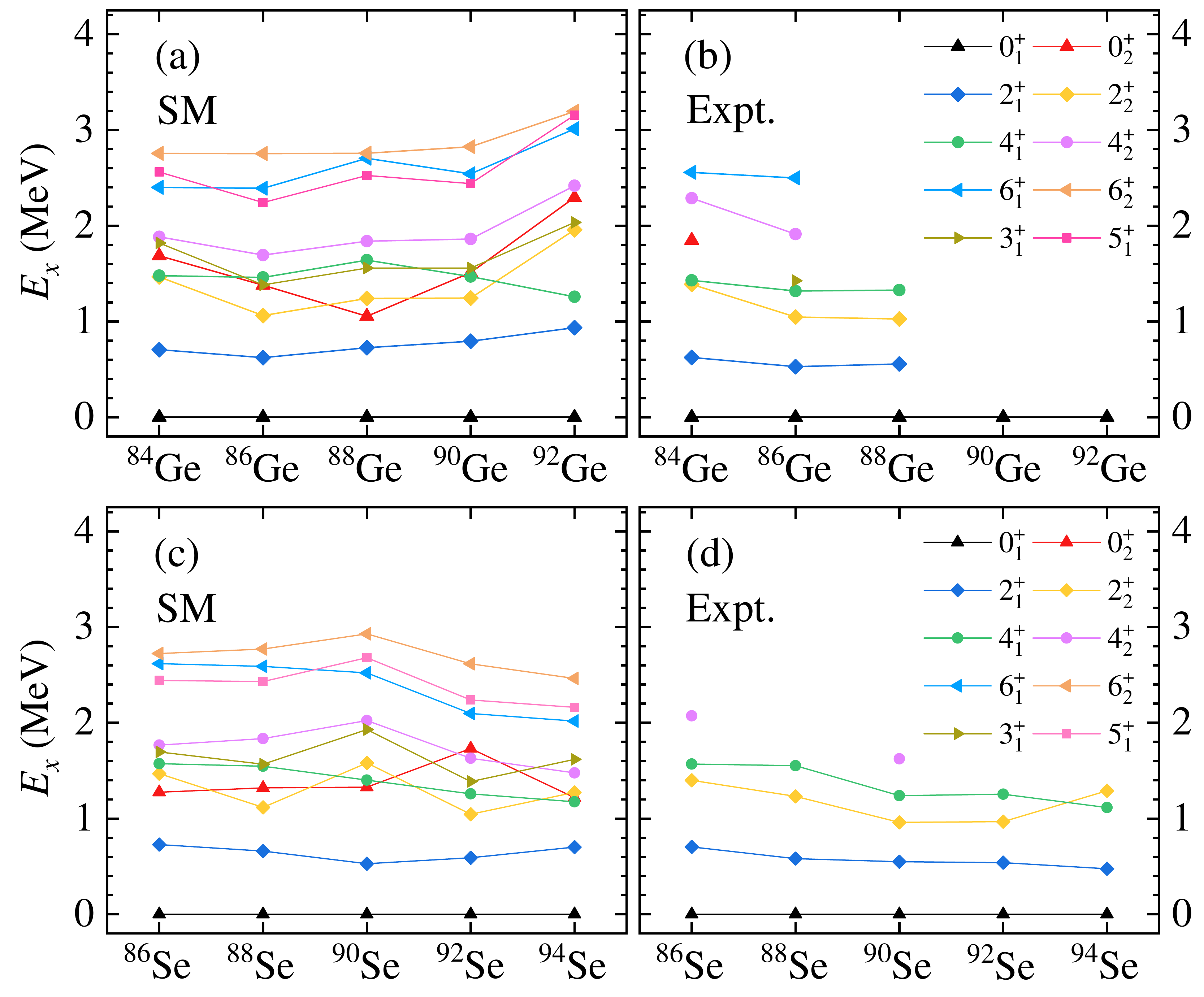}
\caption{\label{F4}{Comparison of experimental and theoretical excitation energy evolution for isotopic chains of Ge ($Z = 32$) and Se ($Z = 34$).}}
\end{figure}

To understand the nuclear shape and collective behaviors in neutron-rich Ge and Se isotopes, we systematically calculate the low-lying states using LSSM calculations.
The  evolutions of excitation energies of $0_{1,2}^+$, $2_{1,2}^+$, $4_{1,2}^+$, $6_{1,2}^+$, $3_{1}^+$, and $5_{1}^+$ states in even-even $^{84-92}$Ge and $^{84-94}$Se isotopes are presented in Fig.~\ref{F4}, along with available experimental data for comparison.

In neutron-rich Ge isotopes, our shell model calculations have successfully depicted the experimental energy trends for states $2^+_1$, $2^+_2$, $4^+_1$, and $6^+_1$, albeit with a few discrepancies. For instance, from $^{86}$Ge to $^{88}$Ge, theoretical calculations indicate an ascending trend for $E(2^+_2)$ and $E(4^+_1)$, contrasting with the relatively flat experimental trends. Simultaneously, theoretical computations also give a slight increase in $E(2^+_1)$ for $^{90,92}$Ge. Moreover, the energy of the $0^+_2$ state displays significant fluctuations with an increase in neutron number, reaching a minimum at $^{88}$Ge. It is challenging to verify whether our calculations of energy trends are accurate for heavier Ge isotopes due to the lack of experimental data for comparison.
The calculated $E(2^+_1)$ and $E(4^+_1)$ values align well with experimental data for the neutron-rich Se isotopes. Furthermore, experimental results indicate that the energy levels of $E(2^+_1)$ and $E(4^+_1)$ continuously decrease as the neutron number increases in Se isotopes, suggesting significant deformations. The trend of $E(4^+_1)$ is consistent with our LSSM calculations. However, the $E(2^+_1)$ level slightly increases when varying from $^{90}$Se to $^{94}$Se. 
The calculated $E(2^+_2)$ also aligns with experimental data, where the $\gamma$-soft band is built above the $2_2^+$ state, except that the calculated $E(2^+_2)$ is higher than $E(4^+_1)$ in $^{90}$Se, contrasting with experimental data.
Overall, good agreement is observed in our shell model calculations compared to experimental data.

Our LSSM calculations suggest the existence of both triaxiality and shape coexistence in neutron-rich Ge and Se isotopes. Moreover, the results for neutron-rich Se isotopes are consistent with self-consistent beyond-mean-field calculations based on the Gogny D1S interaction \cite{PhysRevC.95.041302}. Additionally, our theoretical model predicts the existence of certain states that have not yet been observed experimentally. Discovering these states would represent a significant advancement in future experiments, particularly in determining whether a degree of collectivity, especially triaxiality, is common in nuclei above the magic $^{78}$Ni.

\section{summary}
The present work focuses on advancing the understanding of nuclear structure in neutron-rich isotopes beyond the double magic nucleus $^{78}$Ni. 
A shell model effective interaction is derived  within a model space above $^{78}$Ni inner core for nuclei beyond  $^{78}$Ni. 
We first thoroughly investigate evolutions of states with single-neutron nature above $^{78}$Ni within the $N=51$ isotones, and the many-body configurations are also calculated.
Additionally, the excitation energies of $2_1^+$ states and the corresponding reduced electric quadrupole transition probabilities $B(E2;2^+ \to  0^+)$ for $N=52$ isotones are comprehensively calculated. 
A significant achievement of the calculations is the successful reproduction of the general trend of $B(E2)$ observations in the $N=52$ isotones.
The underlying mechanisms of collectivity within $N=52$ isotones are investigated, evaluating the contributions of \textit{pseudo}-SU(3) and \textit{quasi}-SU(3) symmetries based on the configurations yielded by the LSSM.
The low-lying structures and band characteristics of neutron-rich Ge and Se isotopes are also examined, revealing the theoretical and experimental indications of nonaxial $\gamma$ deformation in certain isotopes like $^{84,86,88}$Ge and $^{86,88,90}$Se. 
Moreover, within these nuclei, both the ground state band and the $\gamma$-soft band are constructed based on the calculated $B(E2)$ transitions.
Furthermore, the present work offers predictions for hitherto unobserved low-lying states in neutron-rich Ge and Se isotopes, providing a valuable theoretical framework for future experimental work. These predictions are poised to significantly enhance our grasp of shell structure evolution and collectivity in the nuclear landscape, particularly in regions abundant with neutrons.


\textit{Acknowledgments.} -- 
This work was supported by the National Key R\&D Program of China under Grant No. 2023YFA1606403; the National Natural Science Foundation of China under Grants No. 12205340, No. 12347106, and No. 12121005; the Gansu Natural Science Foundation under Grants No. 22JR5RA123 and No. 23JRRA614; the Key Research Program of the Chinese Academy of Sciences under Grant No. XDPB15; and the State Key Laboratory of Nuclear Physics and Technology, Peking University under Grant No. NPT2020KFY13.
The numerical calculations in this paper were done using the Hefei advanced computing center.

\bibliographystyle{elsarticle-num_noURL}

\bibliography{Ref}

\end{document}